\documentclass[twocolumn,showpacs,preprintnumbers,amsmath,amssymb]{revtex4}

\newcommand{\msub}[1]{\ensuremath _{\mbox{\scriptsize #1}}} 

\newcommand{\dl}{d \lambda}
\newcommand{\dx}{\Delta x}

\usepackage{graphicx}
\usepackage{dcolumn}
\usepackage{bm}

\begin{document}


\title{Absence of correlations in the QCD Dirac spectrum at high temperature}

\author{Tam\'as G.\ Kov\'acs\footnote{Supported by OTKA Hungarian Science
Fund grants 46925 and 49652 and EU Grant (FP7/2007-2013)/ERC n$^o$208740.}}
\affiliation{%
Department of Physics, University of P\'ecs \\
H-7624 P\'ecs Ifj\'us\'ag u.\ 6., Hungary 
}%


\date{\today}

\begin{abstract}

I propose a simple model of the distribution of the small eigenvalues of the
QCD Dirac operator well above the finite temperature phase transition where
chiral symmetry is restored and the spectral density at zero
vanishes. Assuming the absence of correlations between different regions of
the low lying spectrum I derive analytic formulas for the distribution of the
first two eigenvalues.  I find good agreement with data obtained using the
overlap Dirac operator in quenched $SU(2)$ lattice simulations. This suggests
that if chiral symmetry is restored spectral correlations are not important
and all the statistical properties of the spectrum are encoded in the spectral
density.

\end{abstract}

\pacs{11.15.Ha,12.38.Gc,12.38.Aw,11.30.Rd}
\maketitle

Random Matrix Theory (RMT) has been rather successful in describing some
universal properties of the spectrum of the QCD Dirac operator
\cite{Verbaarschot:2000dy}. Most of the results, however, concern
the chirally broken phase of the theory.  This regime is characterized by a
non-zero density of eigenvalues around zero, which through the Banks-Casher
relation leads to a non-vanishing chiral condensate. Using a single parameter,
the value of the chiral condensate, RMT makes detailed predictions about the
distribution of the smallest Dirac eigenvalues.

Above the finite temperature phase transition, $T_c$, the Dirac spectrum is
much less understood in terms of RMT. Multicritical random matrices with fine
tuning in the action might be able to describe the chirally restored phase
\cite{Akemann:1997wi}. Another possibility is to add a temperature dependent
constant matrix to the usual chiral random matrix to mimic the lowest
Matsubara mode \cite{Jackson:1995nf,Stephanov:1996he}. For comparisons with
lattice staggered data above $T_c$ see Refs.\ \cite{Farchioni:1999ws}.
Recently there has also been renewed interest in a better understanding of the
Dirac spectrum around and above $T_c$ with the hope of clarifying the
connection between the chiral and deconfinement transition
\cite{Gattringer:2006ci}.

In the present paper I propose a model of the low lying Dirac spectrum that
can provide a simple alternative to RMT well above $T_c$ . The main
assumptions I make for the Dirac spectrum around zero are the following:
\begin{enumerate}
\item[(1)] For fixed temperature the spectral density scales with the 
spatial volume. \\[-5mm]
\item[(2)] The spectral density per unit spatial volume is 
\begin{equation}
\rho(\lambda) = C \lambda^\alpha,
  \label{eq:density}
\end{equation}
where $C$ and $\alpha$ are constants. \\[-5mm]
\item[(3)] The number of eigenvalues in any two disjoint intervals are
  independent random variables. 
\end{enumerate}
Using lattice data from quenched $SU(2)$ simulations with the overlap Dirac
operator I directly verify assumptions (1) and (2). Based on (1-3) I derive
analytic formulas for the statistical properties of the first two eigenvalues
in terms of $C, \alpha$ and the spatial volume.  Fitting $C$ and $\alpha$ to
the lattice data, the analytic formulas provide several parameter-free
predictions and I find perfect agreement with the numerical lattice data.

\begin{table}
\caption{\label{tab:confs} The number of configurations $N$ generated for the
different spatial box sizes $L$. $N_{Q=0}$ is the number of configurations
in the trivial topological sector, the one used here.}
\begin{ruledtabular}
\begin{tabular}{|c|rrrrrrrr|}
$L$        &   12  &  14 &  16 &  18 &  20 &  22 &  24 &  32 \\ \hline
$N$        &  900  & 850 & 738 & 400 & 490 & 326 & 376 &  42 \\ \hline
$N_{Q=0}$   &  879  & 821 & 711 & 379 & 451 & 298 & 328 &  29 
\end{tabular}
\end{ruledtabular}
\end{table}

At first I summarize the details of the numerical simulations. The data is
based on quenched simulations of the $SU(2)$ gauge theory with Wilson
plaquette coupling $\beta=2.6$ and time extension $N_t=4$. This corresponds to
a temperature of about $T=2.6T_c$, well above the finite temperature phase
transition.  The spatial size of the box was chosen in the range
$N_s=12-32$, spanning more than an order of magnitude in spatial volume
(see Table \ref{tab:confs}). In eight different spatial volumes I computed the
16 (or 32) eigenvalues of smallest magnitude of the overlap Dirac operator
\cite{Narayanan:ss},
\begin{equation}
 D\msub{ov} = 1 - A \left[A^\dagger A\right]^{-\frac{1}{2}},
 \hspace{1cm} A = 1+s - D_0,
\end{equation}
 where $D_0$ is the Wilson Dirac operator and I use $s=0.4$ that appears 
to be optimal for the condition number of $A^\dagger A$. The spectrum of the
Dirac operator is symmetric with respect to the real axis and I only consider
the eigenvalues with non-negative imaginary parts. The overlap Dirac operator
has an exact chiral symmetry and its spectrum lies along a circle touching the
origin. The eigenvalues of smallest magnitude therefore have only small
real parts and the low end of the spectrum is much like that of the continuum
Dirac operator which has a purely imaginary spectrum. In the analysis I
always use only the imaginary part of the eigenvalues. 

The low-end of the Dirac spectrum is known to depend strongly on the temporal
fermionic boundary condition which is effectively a combination of the
Polyakov loop and the explicitly chosen anti-periodic boundary condition
\cite{Bilgici:2009tx}.  In the quenched $SU(2)$ theory the Polyakov loop
$Z(2)$ symmetry is spontaneously broken above the critical temperature. While
in the negative Polyakov loop sector there is a non-zero density of
eigenvalues around zero (as if chiral symmetry were broken), in the positive
sector there seems to be a gap roughly controlled by the lowest Matsubara
frequency \cite{Kovacs:2008sc}. Although in the quenched theory the two
sectors are equivalent, I will restrict my study to the latter, the
one that survives in the high temperature phase in the presence of dynamical
fermions since they explicitly break the $Z(2)$ symmetry.  The boundary
conditions I use for the fermions are always anti-periodic in the time
direction and periodic in all the spatial directions.

In addition to the small non-zero modes, due to its exact chiral symmetry, the
overlap operator can also have exact zero modes, regardless of
the boundary condition or the Polyakov loop sector. 
I use only configurations
belonging to the trivial topological sector which is the most abundant in
these simulations (see Table \ref{tab:confs}).

\begin{figure}
\vspace{1cm}
\includegraphics[width=\columnwidth,keepaspectratio]{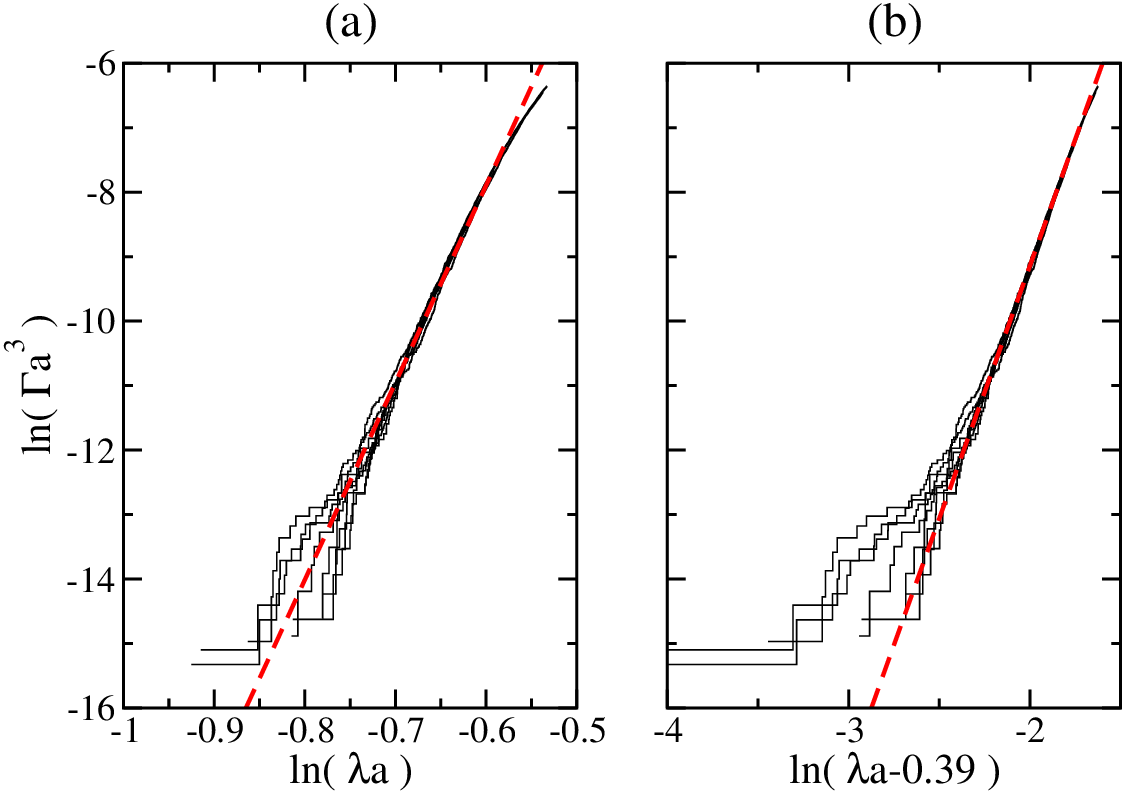}
\caption{\label{fig:Gammafit} The cumulative spectral density scaled by 
the 3-volume for all eight 3-volumes. The dashed line is the analytic
prediction using the parameters fitted for the average smallest eigenvalue.
The difference between Fig.\ (a) and (b) is that while in the former we assume
no gap for the fitting, in the latter a gap $G=0.39$ is assumed.}
\end{figure}

{\em (1) Scaling of the spectral density---} In Fig.\ \ref{fig:Gammafit} I
plot the cumulative spectral density normalized by the spatial volume,
    %
\begin{equation}
\Gamma(\lambda)=\int_0^\lambda dx \, \rho(x),
\end{equation} 
for different spatial box sizes.
$\Gamma(\lambda)$ is the average number of eigenvalues per unit volume smaller
than $\lambda$. The densities in the different volumes agree perfectly except
for the very low end where the data is sparse and statistical errors are
large.  The scaling of the spectral density with the 3-volume is a rather
non-trivial property of the interacting theory. For
free fermions the level spacing is inversely proportional to the linear box
size since the Dirac equation is first order. That would imply a spectral
density scaling with the linear size, not the volume.

{\em (2) Form of the spectral density---} As can be seen in the log-log plot
of $\Gamma$, the integrated spectral density versus $\lambda$
(Fig.\ \ref{fig:Gammafit}a), the data is well described by the simple power
law of Eq.\ (\ref{eq:density}) therefore I will assume that in what follows.

I emphasize that for the main result of the paper the particular form of the
spectral density is immaterial. The whole forthcoming analysis can be carried
out with any simple analytic form of the spectral density that describes the
data accurately.  In particular there are also expectations based partly on
RMT \cite{Jackson:1995nf} and lattice data \cite{Gattringer:2002tg} that at
high temperature, well above $T_c$ the spectrum develops a gap $G$ around
zero.  In that case the spectral density would be expected to start off from
$G$ as
\begin{equation}
 \rho(\lambda) = C (\lambda-G )^\alpha.
     \label{eq:rhogap}
\end{equation}
Later I will also discuss this possibility. Potentially the accumulation of
small non-zero eigenvalues seen in \cite{Edwards:1999zm} might also complicate
the picture, but the temperature used here is high enough that these modes are
completely absent, at least with the present statistics.

{\em (3) No spectral correlations---} i.e.\ the number of eigenvalues in any
two disjoint intervals are independent random variables. This is the most
non-trivial of the three assumptions and the demonstration of its validity is
the main result of the paper. I will show that the resulting description fits
the numerical data rather accurately.  The absence of correlations between
different regions of the spectrum immensely simplifies any model since it
means that all probabilistic information about the spectrum is encoded in the
spectral density. All the properties of the low eigenvalues follow from a
simple few-parameter description of the spectral density like
Eq.\ (\ref{eq:density}) or (\ref{eq:rhogap}).

It is important to note that the absence of correlations between different
regions of the spectrum is not related in any simple way to the correlations
between individual eigenvalues that is often considered in random matrix
models \cite{Akemann:2003tv}.  In particular it does not imply the absence of
correlations between individual eigenvalues.

{\em Distribution of the smallest eigenvalues---} I now compute
the distribution of the smallest two eigenvalues from 
assumptions (1-3). From the definition of the spectral density
$V\rho(\lambda)\dl$ is the average number of eigenvalues in spatial volume $V$
in an interval of length $\dl$ centered around $\lambda$. If at fixed volume
the length of the interval $\dl\rightarrow~0$ then the probability of having
one eigenvalue in the interval is $V\rho(\lambda)\dl$ and the probability of
no eigenvalue is $1-V\rho(\lambda)\dl$. This is because the probability of
having more than one eigenvalue becomes negligible.

Let us first compute the probability $P_{no}(\lambda_1,\lambda_2)$ that there
is no eigenvalue in the interval $[\lambda_1,\lambda_2]$. Since 
correlations between the subintervals are ignored the probability of having
no eigenvalue in an interval can be written as the product of probabilities
of having no eigenvalue in any subinterval of its decomposition into 
small subintervals.
\begin{eqnarray}
  P_{no}(\lambda_1,\lambda_2) = \lim_{\dx\rightarrow 0} 
                   (1-V\rho(x_1) \dx) \; (1-V\rho(x_2) \dx)\;...  
    \nonumber \\
                   (1-V\rho(x_N) \dx), 
\end{eqnarray}
where $x_1=\lambda_1$, $x_N=\lambda_2$ and $x_{k+1}-x_k=\dx$. Expanding the
product the terms can be organized in powers of $\dx$ and in the limit
$\dx\rightarrow 0$ the order $n$ term goes to the following $n$-fold integral;
\begin{eqnarray}
 S_n = (-V)^n \int_{\lambda_1}^{\lambda_2} dx_1 \rho(x_1)
              \int_{x_1}^{\lambda_2} dx_2 \rho(x_2) \; ... \;
   \nonumber \\
              \int_{x_{n-1}}^{\lambda_2} dx_n \rho(x_n).
\end{eqnarray}
Substituting the spectral density in Eq.\ (\ref{eq:density}) for $\rho$ the
integrations can be explicitly carried out resulting in 
\begin{equation}
 S_n = \frac{(-CV)^n}{n!(\alpha+1)^n} 
       \left(  \lambda_2^{\alpha+1} -  \lambda_1^{\alpha+1} \right)^n.
\end{equation}
Finally the summation over $n$ is easily done giving
\begin{equation}
P_{no}(\lambda_1,\lambda_2) = \exp \left( 
    -\frac{CV}{\alpha+1}( \lambda_2^{\alpha+1} - \lambda_1^{\alpha+1} )
  \right)
\end{equation}
for the probability of no eigenvalue in the interval $[\lambda_1,\lambda_2]$.

Using again assumption (3) the probability of having the smallest eigenvalue
around $\lambda$ is the product of two probabilities: having no eigenvalue in
$[0,\lambda]$ and having one eigenvalue in $[\lambda,\lambda+d\lambda]$. Thus
the probability density of the smallest eigenvalue is

\begin{equation}
 \rho_1(\lambda) = \exp\left( -\frac{CV}{\alpha+1} \lambda^{\alpha+1} \right)
                   CV\lambda^\alpha.
    \label{eq:rho1}
\end{equation}
The average smallest eigenvalue is easily calculated from this as 
\begin{equation}
 \langle \lambda_1 \rangle = \int_0^\infty \rho_1(x) x \; dx =
 (CV\mu)^{-\mu} \; \Gamma(1+\mu),
     \label{eq:avsmallest}
\end{equation}
where I introduced the notation $\mu=(1+\alpha)^{-1}$.

The distribution and average of the second smallest eigenvalue can be
calculated in a similar fashion. Again the probability of having the
second eigenvalue around $\lambda$ can be decomposed as a product: having the
smallest eigenvalue at $x$, having no eigenvalues between $x$ and $\lambda$
and having an eigenvalue around $\lambda$. Since $x$ (the occurrence of the
first eigenvalue) can be anywhere between 0 and $\lambda$, $x$ has to be 
integrated and the probability density of the second eigenvalue is 
\begin{eqnarray}
 \rho_2(\lambda) & = & \int_0^\lambda dx \; \rho_1(x) \; P_{no}(x,\lambda) \;
 \rho(\lambda)   \nonumber \\
  & = & \frac{C^2V^2}{\alpha+1} \exp\left( -\frac{CV}{\alpha+1} 
      \lambda^{\alpha+1} \right) \, \lambda^{2\alpha+1}.
     \label{eq:rho2} 
\end{eqnarray}
Finally we obtain 
\begin{equation}
 \langle \lambda_2 \rangle = (CV\mu)^{-\mu} \; \Gamma(2+\mu),
     \label{eq:av2ndev}
\end{equation}
for the average second eigenvalue. The procedure can be easily continued for
higher eigenvalues.

\begin{figure}
\vspace{6mm}
\includegraphics[width=\columnwidth,keepaspectratio]{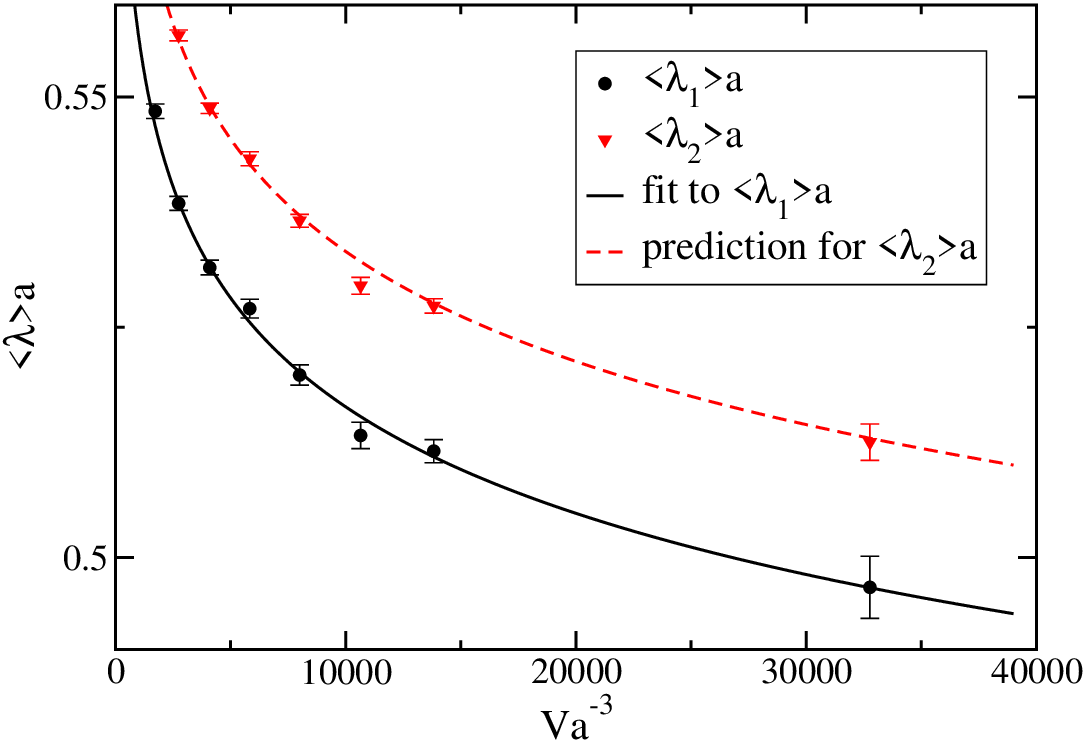}
\caption{\label{fig:fitsmallest} The average smallest and second smallest
eigenvalue as a function of the spatial volume in lattice units. The solid
curve is a two-parameter fit of the form Eq.\ (\ref{eq:avsmallest}). The
dashed curve is the fitted parameter-free prediction for the 
average second smallest eigenvalue.}
\end{figure}

{\em Numerical results and tests---} In the remainder I compare
the analytic formulas and the data. There are many different ways
to proceed; one has to fit the two parameters $C$ and $\alpha$ in the spectral
density using any of the analytic formulas above and then the rest of the
formulas are predictions that can be tested against the data.  I choose to use
Eq.\ (\ref{eq:avsmallest}), the scaling of the average smallest eigenvalue
with the 3-volume $V$ for the fit. Using a fit range of $V=14^3-32^3$ produces
a good fit with $\chi^2$/d.o.f.$=0.89$. The parameters obtained are
$C=1.07(82)\times 10^6$ and $\mu=\frac{1}{\alpha+1}=0.0327(12)$. In
Fig.\ \ref{fig:fitsmallest} I plotted the average smallest eigenvalue as a
function of the 3-volume along with the fit. Also in this Figure I present the
average second eigenvalue and the analytic formula of Eq.\ (\ref{eq:av2ndev})
using the above parameters. Notice that this is now a prediction of the model
with no further parameters to be fitted and it agrees quite well with the
data.

\begin{figure}
\includegraphics[width=\columnwidth,keepaspectratio]{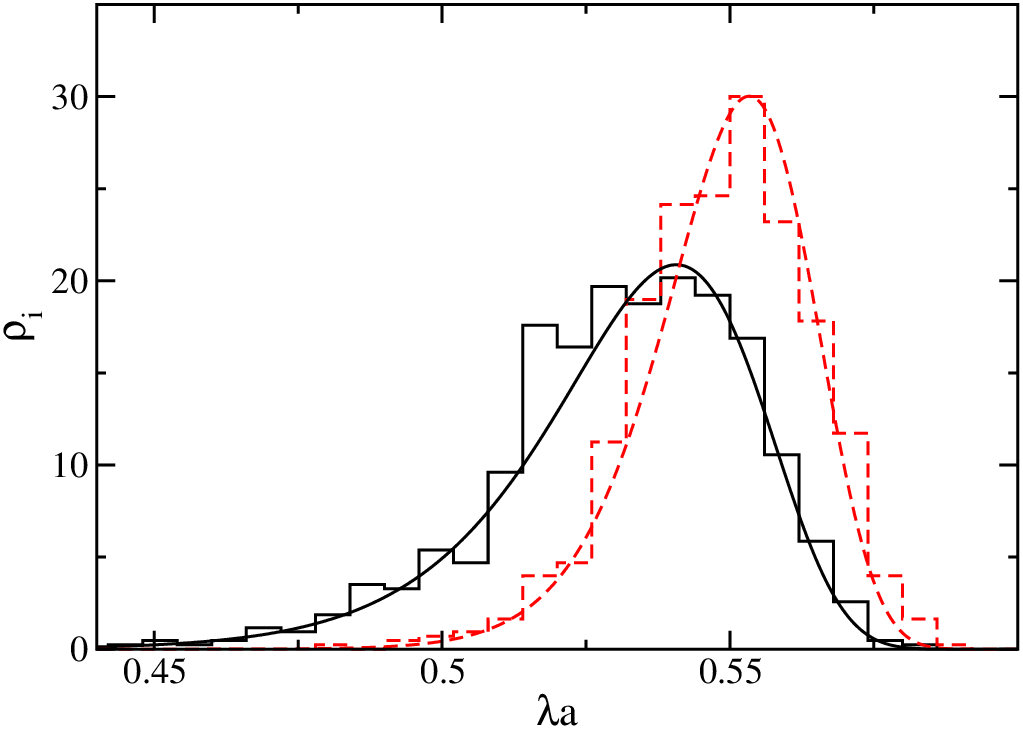}
\caption{\label{fig:smallestdist} The distribution of the smallest and 
second smallest eigenvalues in a spatial volume of $V=16^3$. The solid 
lines indicate the analytic prediction of the model (Eqs.\ 
(\ref{eq:rho1}) and (\ref{eq:rho2})).}
\end{figure}

A more detailed prediction is the distribution of the smallest and second
smallest eigenvalue for different spatial volumes. As an illustration, Fig.\
\ref{fig:smallestdist} shows this in a spatial volume of $V=16^3$ and again
there is good agreement with the data. The picture is similar for the other 
volumes down to $V=14^3$ below which even the distribution of the 
smallest eigenvalue starts to be concentrated above the upper limit of the 
validity of the simple power-law of Eq.\ (\ref{eq:density}) for the spectral 
density. 

{\em Is there a spectral gap?---} Coming back to the question of whether there
is a spectral gap, the derivation of the distributions can be easily 
generalized for the spectral density of Eq.\ (\ref{eq:rhogap}) yielding
\begin{equation}
 \langle \lambda_1 \rangle = 
 (CV\mu)^{-\mu} \; \Gamma(1+\mu) \; + \; G
\end{equation}
for the average smallest eigenvalue.
Unfortunately a three-parameter fit---including $G$---to the average
smallest eigenvalue is not very meaningful since in a wide range of $G$ almost
equally good fits can be found for $C$ and $\alpha$. We can, however, assume
that $G$ is somewhat smaller than the smallest observed eigenvalue.  In our
simulation data for the eight different spatial volumes comprising altogether
more than 4000 configurations the smallest eigenvalue was 0.3965. If a gap of
$G=0.39$ is assumed, the fit yields $C=5.4(3.6)\times 10^3$ and
$\mu=0.1275(52)$ with $\chi^2/$d.o.f.$=1.03$. Comparing the predicted
cumulative spectral density with the data on a log-log plot 
(Figs.\ \ref{fig:Gammafit}a and \ref{fig:Gammafit}b)  shows that the cumulative
spectral density definitely favors $G=0$ over $G=0.39$, however, a much
smaller but still non-zero value of the gap cannot be ruled out based on the
present data.

{\em Discussion---} We proposed a description of the low-lying Dirac spectrum
at high temperature based on the assumption that there are no correlations
between different regions in the spectrum.  Our model provides detailed
analytic predictions for the distribution of the low-lying Dirac eigenvalues
and perfect agreement is found with numerical lattice data.  The absence of
correlations is a peculiar property of the chirally symmetric phase and is not
expected to hold if chiral symmetry is broken.  It would be interesting on the
one hand to get a more fundamental understanding of this property and on the
other hand to see whether it is shared by a wider class of chirally symmetric
systems. A good testing ground could be the Schwinger model with two flavors
of massless fermions \cite{Bietenholz:2009jn}.  The model proposed here might
eventually provide a full description of the Dirac spectrum in the chirally
restored phase, just like Random Matrix Theory in the chirally broken
phase. Together with RMT it might also open the possibility to a better
understanding of the nature of the chiral phase transition of QCD and even
the conformal phase of QCD-like theories with more fermionic degrees of
freedom \cite{DeGrand:2009et}.


\begin{thebibliography}{99}

\bibitem{Verbaarschot:2000dy}
  J.~J.~M.~Verbaarschot and T.~Wettig,
  Ann.\ Rev.\ Nucl.\ Part.\ Sci.\  {\bf 50}, 343 (2000)
  [arXiv:hep-ph/0003017].

\bibitem{Akemann:1997wi}
  G.~Akemann, P.~H.~Damgaard, U.~Magnea and S.~M.~Nishigaki,
  Nucl.\ Phys.\  B {\bf 519}, 682 (1998)
  [arXiv:hep-th/9712006].

\bibitem{Jackson:1995nf}
  A.~D.~Jackson and J.~J.~M.~Verbaarschot,
  Phys.\ Rev.\  D {\bf 53}, 7223 (1996)
  [arXiv:hep-ph/9509324].

\bibitem{Stephanov:1996he}
  M.~A.~Stephanov,
  Phys.\ Lett.\  B {\bf 375}, 249 (1996)
  [arXiv:hep-lat/9601001];
  M.~A.~Nowak, G.~Papp and I.~Zahed,
  Phys.\ Lett.\  B {\bf 389}, 341 (1996)
  [arXiv:hep-ph/9604235].

\bibitem{Farchioni:1999ws}
  F.~Farchioni, P.~de Forcrand, I.~Hip, C.~B.~Lang and K.~Splittorff,
  Phys.\ Rev.\  D {\bf 62}, 014503 (2000)
  [arXiv:hep-lat/9912004];
  P.~H.~Damgaard, U.~M.~Heller, R.~Niclasen and K.~Rummukainen,
  Nucl.\ Phys.\  B {\bf 583}, 347 (2000)
  [arXiv:hep-lat/0003021].

\bibitem{Gattringer:2006ci}
  C.~Gattringer,
  Phys.\ Rev.\ Lett.\  {\bf 97}, 032003 (2006)
  [arXiv:hep-lat/0605018];
  F.~Bruckmann, C.~Gattringer and C.~Hagen,
  Phys.\ Lett.\  B {\bf 647}, 56 (2007)
  [arXiv:hep-lat/0612020].

\bibitem{Gattringer:2002tg}
  C.~Gattringer and S.~Schaefer,
  Nucl.\ Phys.\  B {\bf 654}, 30 (2003)
  [arXiv:hep-lat/0212029].

\bibitem{Narayanan:ss}
R.~Narayanan and H.~Neuberger,
Phys.\ Rev.\ Lett.\  {\bf 71} (1993) 3251
[arXiv:hep-lat/9308011].
Nucl.\ Phys.\ B {\bf 412}, 574 (1994)
[arXiv:hep-lat/9307006];
Nucl.\ Phys.\ B {\bf 443}, 305 (1995)
[arXiv:hep-th/9411108].

\bibitem{Bilgici:2009tx}
  E.~Bilgici, F.~Bruckmann, J.~Danzer, C.~Gattringer, C.~Hagen, 
  E.~M.~Ilgenfritz and A.~Maas,
  arXiv:0906.3957 [hep-lat].

\bibitem{Edwards:1999zm}
  R.~G.~Edwards, U.~M.~Heller, J.~E.~Kiskis and R.~Narayanan,
  Phys.\ Rev.\  D {\bf 61}, 074504 (2000)
  [arXiv:hep-lat/9910041].

\bibitem{Kovacs:2008sc}
  T.~G.~Kovacs,
  PoS {\bf LATTICE2008}, 198 (2008)
  [arXiv:0810.4763 [hep-lat]].

\bibitem{Akemann:2003tv}
  G.~Akemann and P.~H.~Damgaard,
  Phys.\ Lett.\  B {\bf 583}, 199 (2004)
  [arXiv:hep-th/0311171].

\bibitem{DeGrand:2009et}
  T.~DeGrand,
  arXiv:0906.4543 [hep-lat];
        %
  Z.~Fodor, K.~Holland, J.~Kuti, D.~Nogradi and C.~Schroeder,
  arXiv:0908.2466 [hep-lat].

\bibitem{Bietenholz:2009jn}
  W.~Bietenholz and I.~Hip,
  arXiv:0909.2241 [hep-lat].



\end{thebibliography}
\end{document}